\documentclass[sigconf]{acmart}
\hypersetup{draft}

\usepackage[T1]{fontenc}
\usepackage{afterpage}
\usepackage{listings}
\usepackage{xcolor}
\usepackage{graphicx}
\usepackage{float}
\usepackage{caption}
\usepackage{subcaption}
\usepackage{placeins}
\usepackage{url}
\usepackage{epstopdf}
\usepackage{flushend}
\usepackage{booktabs}
\usepackage{footnote}
\usepackage[htt]{hyphenat}
\makesavenoteenv{table}
\makesavenoteenv{tabular}

\usepackage[acronyms,nonumberlist,nopostdot,nomain,nogroupskip]{glossaries}

\usepackage{tabularx}

\usepackage{paralist}

\usepackage{enumerate}

\newacronym{3gpp}{3GPP}{3rd Generation Partnership Project}
\newacronym{adc}{ADC}{Analog to Digital Converter}
\newacronym{5g}{5G}{5th generation}
\newacronym{aimd}{AIMD}{Additive Increase Multiplicative Decrease}
\newacronym{am}{AM}{Acknowledged Mode}
\newacronym{amc}{AMC}{Adaptive Modulation and Coding}
\newacronym{aqm}{AQM}{Active Queue Management}
\newacronym{awgn}{AGWN}{Additive White Gaussian Noise}
\newacronym{balia}{BALIA}{Balanced Link Adaptation}
\newacronym{bdp}{BDP}{Bandwidth-Delay Product}
\newacronym{bf}{BF}{Beamforming}
\newacronym{cc}{CC}{Congestion Control}
\newacronym{cdf}{CDF}{Cumulative Distribution Function}
\newacronym{cn}{CN}{Core Network}
\newacronym{cqi}{CQI}{Channel Quality Information}
\newacronym{cp}{CP}{Control Plane}
\newacronym{csirs}{CSI-RS}{Channel State Information - Reference Signal}
\newacronym{dc}{DC}{Dual Connectivity}
\newacronym{dce}{DCE}{Direct Code Execution}
\newacronym{dci}{DCI}{Downlink Control Information}
\newacronym{dl}{DL}{Downlink}
\newacronym{dmr}{DMR}{Deadline Miss Ratio}
\newacronym{dmrs}{DMRS}{DeModulation Reference Signal}
\newacronym{e2e}{E2E}{end-to-end}
\newacronym{ecn}{ECN}{Explicit Congestion Notification}
\newacronym{edf}{EDF}{Earliest Deadline First}
\newacronym{enb}{eNB}{evolved Node Base}
\newacronym{epc}{EPC}{Evolved Packet Core}
\newacronym{es}{ES}{Edge Server}
\newacronym{fdma}{FDMA}{Frequency Division Multiple Access}
\newacronym{fdd}{FDD}{Frequency Division Duplexing}
\newacronym[firstplural=Radio Access Technologies (RATs)]{rat}{RAT}{Radio Access Technology}
\newacronym{fs}{FS}{Fast Switching}
\newacronym{ftp}{FTP}{File Transfer Protocol}
\newacronym{gnb}{gNB}{Next Generation Node Base}
\newacronym{harq}{HARQ}{Hybrid Automatic Repeat reQuest}
\newacronym{hetnet}{HetNet}{Heterogeneous Network}
\newacronym{hh}{HH}{Hard Handover}
\newacronym{hol}{HOL}{Head-of-Line}
\newacronym{ia}{IA}{Initial Access}
\newacronym{ieee}{IEEE}{Institute of Electrical and Electronics Engineers}
\newacronym{imt}{IMT}{International Mobile Telecommunication}
\newacronym{iot}{IoT}{Internet of Things}
\newacronym{ldpc}{LDPC}{Low-Density Parity Check}
\newacronym{los}{LOS}{Line-of-Sight}
\newacronym{lte}{LTE}{Long Term Evolution}
\newacronym{m2m}{M2M}{Machine to Machine}
\newacronym{mac}{MAC}{Medium Access Control}
\newacronym{mc}{MC}{Multi-Connectivity}
\newacronym{mcs}{MCS}{Modulation and Coding Scheme}
\newacronym{mec}{MEC}{Mobile Edge Cloud}
\newacronym{mi}{MI}{Mutual Information}
\newacronym{mimo}{MIMO}{Multiple Input, Multiple Output}
\newacronym{mmwave}{mmWave}{millimeter wave}
\newacronym{mptcp}{MPTCP}{Multipath TCP}
\newacronym{mr}{MR}{Maximum Rate}
\newacronym{mss}{MSS}{Maximum Segment Size}
\newacronym{mtd}{MTD}{Machine-Type Device}
\newacronym{mtu}{MTU}{Maximum Transmission Unit}
\newacronym{nfv}{NFV}{Network Function Virtualization}
\newacronym{nlos}{NLOS}{Non-Line-of-Sight}
\newacronym{nlosv}{NLOSv}{Vehicle Non-Line-of-Sight}
\newacronym{nr}{NR}{New Radio}
\newacronym{ofdm}{OFDM}{Orthogonal Frequency Division Multiplexing}
\newacronym{pdcch}{PDCCH}{Physical Downlonk Control Channel}
\newacronym{pdcp}{PDCP}{Packet Data Convergence Protocol}
\newacronym{pdsch}{PDSCH}{Physical Downlink Shared Channel}
\newacronym{pdu}{PDU}{Packet Data Unit}
\newacronym{pf}{PF}{Proportional Fair}
\newacronym{pgw}{PGW}{Packet Gateway}
\newacronym{phy}{PHY}{Physical}
\newacronym{pbch}{PBCH}{Physical Broadcast Channel}
\newacronym[plural=\gls{mme}s,firstplural=Mobility Management Entities (MMEs)]{mme}{MME}{Mobility Management Entity}
\newacronym{prb}{PRB}{Physical Resource Block}
\newacronym{pss}{PSS}{Primary Synchronization Signal}
\newacronym{pscch}{PSCCH}{Physical Sidelink Control Channel}
\newacronym{pucch}{PUCCH}{Physical Uplink Control Channel}
\newacronym{pusch}{PUSCH}{Physical Uplink Shared Channel}
\newacronym{rach}{RACH}{Random Access Channel}
\newacronym{ran}{RAN}{Radio Access Network}
\newacronym{red}{RED}{Random Early Detection}
\newacronym{rf}{RF}{Radio Frequency}
\newacronym{rlc}{RLC}{Radio Link Control}
\newacronym{rlf}{RLF}{Radio Link Failure}
\newacronym{rrc}{RRC}{Radio Resource Control}
\newacronym{rrm}{RRM}{Radio Resource Management}
\newacronym{rr}{RR}{Round Robin}
\newacronym{rs}{RS}{Remote Server}
\newacronym{rsrp}{RSRP}{Reference Signal Received Power}
\newacronym{rss}{RSS}{Received Signal Strength}
\newacronym{rtt}{RTT}{Round Trip Time}
\newacronym{rw}{RW}{Receive Window}
\newacronym{rx}{RX}{Receiver}
\newacronym{sa}{SA}{standalone}
\newacronym{sack}{SACK}{Selective Acknowledgment}
\newacronym{sap}{SAP}{Service Access Point}
\newacronym{sc}{SC}{Single Carrier}
\newacronym{sch}{SCH}{Secondary Cell Handover}
\newacronym{scoot}{SCOOT}{Split Cycle Offset Optimization Technique}
\newacronym{sdma}{SDMA}{Spatial Division Multiple Access}
\newacronym{sinr}{SINR}{Signal to Interference plus Noise Ratio}
\newacronym{sl}{SL}{Sidelink}
\newacronym{sm}{SM}{Saturation Mode}
\newacronym{snr}{SNR}{Signal-to-Noise-Ratio}
\newacronym{son}{SON}{Self-Organizing Network}
\newacronym{ss}{SS}{Synchronization Signal}
\newacronym{srs}{SRS}{Sounding Reference Signal}
\newacronym{sss}{SSS}{Secondary Synchronization Signal}
\newacronym{tb}{TB}{Transport Block}
\newacronym{tcp}{TCP}{Transmission Control Protocol}
\newacronym{tdd}{TDD}{Time Division Duplexing}
\newacronym{tdma}{TDMA}{Time Division Multiple Access}
\newacronym{tfl}{TfL}{Transport for London}
\newacronym{tm}{TM}{Transparent Mode}
\newacronym{trp}{TRP}{Transmitter Receiver Pair}
\newacronym{tti}{TTI}{Transmission Time Interval}
\newacronym{ttt}{TTT}{Time-to-Trigger}
\newacronym{tx}{TX}{Transmitter}
\newacronym{ue}{UE}{User Equipment}
\newacronym{ul}{UL}{Uplink}
\newacronym{uml}{UML}{Unified Modeling Language}
\newacronym{um}{UM}{Unacknowledged Mode}
\newacronym{utc}{UTC}{Urban Traffic Control}
\newacronym{vm}{VM}{Virtual Machine}
\newacronym{rsrq}{RSRQ}{Reference Signal Received Quality}
\newacronym{rssi}{RSSI}{Received Signal Strength Indicator}
\newacronym{crs}{CRS}{Cell Reference Signal}
\newacronym{nsa}{NSA}{Non Stand Alone}
\newacronym{mrdc}{MR-DC}{Multi \gls{rat} \gls{dc}}
\newacronym{endc}{EN-DC}{E-UTRAN-\gls{nr} \gls{dc}}
\newacronym{5gc}{5GC}{5G Core}
\newacronym{si}{SI}{Study Item}
\newacronym{iab}{IAB}{Integrated Access and Backhaul}
\newacronym{wf}{WF}{Wired-first}
\newacronym{hqf}{HQF}{Highest-quality-first}
\newacronym{pa}{PA}{Position-aware}
\newacronym{mlr}{MLR}{Maximum-local-rate}
\newacronym{wbf}{WBF}{Wired Bias Function}
\newacronym{mib}{MIB}{Master Information Block}
\newacronym{sib}{SIB}{Secondary Information Block}
\newacronym{rnti}{RNTI}{Radio Network Temporary Identifier}
\newacronym{dft}{DFT}{Discrete Fourier Transform}
\newacronym{kpi}{KPI}{Key Performance Indicator}
\newacronym{ppp}{PPP}{Poisson Point Process}
\newacronym{v2v}{V2V}{Vehicle-to-Vehicle}
\newacronym{wave}{WAVE}{Wireless Access in Vehicular Environments}
\newacronym{udp}{UDP}{User Datagram Protocol}
\newacronym{upa}{UPA}{Uniform Planar Array}
\newacronym{fec}{FEC}{Forward Error Correction}
\newacronym{v2x}{V2X}{Vehicle-To-Everything}
\newacronym{psfch}{PSFCH}{Physical Sidelink Feedback Channel}
\newacronym{pssch}{PSSCH}{Physical Sidelink Shared Channel}
\newacronym{csma}{CSMA}{Carrier Sense Multiple Access}
\newacronym{v2n}{V2N}{Vehicle-to-Network}
\newacronym{wlan}{WLAN}{Wireless Local Area Network}
\newacronym{cav}{CAV}{Connected and Autonomous Vehicle}
\newacronym{v2i}{V2I}{Vehicle-to-Infrastructure}
\newacronym{d2d}{D2D}{Device-to-Device}
\newacronym{c-its}{C-ITS}{Connected Intelligent Transportation System}
\newacronym{fr2}{FR2}{Frequency Range 2}
\newacronym{bs}{BS}{Base Station}
\newacronym{sdu}{SDU}{Service Data Unit}
\newacronym{csi}{CSI}{Channel State Information}
\newacronym{scs}{SCS}{Subcarrier Spacing}

\usepackage{tikz}
\usepackage{pgfplots}
\pgfplotsset{compat=newest} 
\pgfplotsset{plot coordinates/math parser=false} 
\newlength\fheight
\newlength\fwidth
\usetikzlibrary{plotmarks,patterns,decorations.pathreplacing,backgrounds,calc,arrows,arrows.meta,spy,matrix,automata,positioning}
\usepgfplotslibrary{patchplots,groupplots}
\usepackage{tikzscale}

\tikzstyle{startstop} = [rectangle, rounded corners, minimum width=2cm, minimum height=0.5cm,text centered, draw=black]
\tikzstyle{io} = [trapezium, trapezium left angle=70, trapezium right angle=110, minimum width=3cm, minimum height=1cm, text centered, draw=black]
\tikzstyle{process} = [rectangle, minimum width=2cm, minimum height=0.5cm, text centered, draw=black, align=center]
\tikzstyle{decision} = [ellipse, minimum width=2cm, minimum height=1cm, text centered, draw=black]
\tikzstyle{arrow} = [thick,<->,>=stealth]
\tikzstyle{line} = [thick,>=stealth]
\tikzstyle{darrow} = [thick,<->,>=stealth,dashed]
\tikzstyle{sarrow} = [thick,->,>=stealth]
\tikzstyle{larrow} = [line width=0.1mm,dashdotted,<->,>=stealth]

\definecolor{SchoolColor}{RGB}{0.71, 0, 0.106}%181,0,27} unipd red
\definecolor{chaptergrey}{rgb}{0.61, 0, 0.09} % dialed back a little
\definecolor{midgrey}{rgb}{0.4, 0.4, 0.4}
\definecolor{chaptergreen}{rgb}{0.09, 0.612, 0}
\definecolor{chapterpurple}{rgb}{0.522, 0, 0.612}
\definecolor{chapterlightgreen}{rgb}{0, 0.612, 0.522}

\makeatletter
\def\grd@save@target#1{%
  \def\grd@target{#1}}
\def\grd@save@start#1{%
  \def\grd@start{#1}}
\tikzset{
  grid with coordinates/.style={
    to path={%
      \pgfextra{%
        \edef\grd@@target{(\tikztotarget)}%
        \tikz@scan@one@point\grd@save@target\grd@@target\relax
        \edef\grd@@start{(\tikztostart)}%
        \tikz@scan@one@point\grd@save@start\grd@@start\relax
        \draw[minor help lines] (\tikztostart) grid (\tikztotarget);
        \draw[major help lines] (\tikztostart) grid (\tikztotarget);
        \grd@start
        \pgfmathsetmacro{\grd@xa}{\the\pgf@x/1cm}
        \pgfmathsetmacro{\grd@ya}{\the\pgf@y/1cm}
        \grd@target
        \pgfmathsetmacro{\grd@xb}{\the\pgf@x/1cm}
        \pgfmathsetmacro{\grd@yb}{\the\pgf@y/1cm}
        \pgfmathsetmacro{\grd@xc}{\grd@xa + \pgfkeysvalueof{/tikz/grid with coordinates/major step x}}
        \pgfmathsetmacro{\grd@yc}{\grd@ya + \pgfkeysvalueof{/tikz/grid with coordinates/major step y}}
        \foreach \x in {\grd@xa,\grd@xc,...,\grd@xb}
        \node[anchor=north] at (\x,\grd@ya) {\pgfmathprintnumber{\x}};
        \foreach \y in {\grd@ya,\grd@yc,...,\grd@yb}
        \node[anchor=east] at (\grd@xa,\y) {\pgfmathprintnumber{\y}};
      }
    }
  },
  minor help lines/.style={
    help lines,
    gray,
    line cap =round,
    xstep=\pgfkeysvalueof{/tikz/grid with coordinates/minor step x},
    ystep=\pgfkeysvalueof{/tikz/grid with coordinates/minor step y}
  },
  major help lines/.style={
    help lines,
    line cap =round,
    line width=\pgfkeysvalueof{/tikz/grid with coordinates/major line width},
    xstep=\pgfkeysvalueof{/tikz/grid with coordinates/major step x},
    ystep=\pgfkeysvalueof{/tikz/grid with coordinates/major step y}
  },
  grid with coordinates/.cd,
  minor step x/.initial=.5,
  minor step y/.initial=.2,
  major step x/.initial=1,
  major step y/.initial=1,
  major line width/.initial=1pt,
}
\makeatother

\newcommand{\millicar}[0]{MilliCar}

\begin{document}
% reduce space before and after eq

%\setlength{\abovedisplayskip}{2pt}
%\setlength{\belowdisplayskip}{2pt}
% space after Figure captions
% \setlength{\belowcaptionskip}{-0.4cm}
%\setlength{\headsep}{0.05in}
%\setlength{\topskip}{-10mm}

%align the bottom of the pages
\flushbottom
\setlength{\parskip}{0ex plus0.1ex}
\addtolength{\skip\footins}{-0.2pc plus 40pt}

\title{\millicar{} - An ns-3 Module for mmWave NR V2X Networks}
%\numberofauthors{3} %  in this sample file, there are a *total*

% \author{Tommaso Zugno}
% \affiliation{Department of Information Engineering,\\University of Padova,\\Padova, Italy}
% \email{zugnotom@dei.unipd.it}

% \author{Michele Polese}
% \affiliation{Department of Information Engineering,\\University of Padova,\\Padova, Italy}
% \email{polesemi@dei.unipd.it}

% \author{Michele Zorzi}
% \affiliation{Department of Information Engineering,\\University of Padova,\\Padova, Italy}
% \email{zorzi@dei.unipd.it}

\author{\texorpdfstring{Matteo Drago, Tommaso Zugno, Michele Polese, Marco Giordani, Michele Zorzi\\
\small Department of Information Engineering, University of Padova, Padova, Italy \\
\small e-mail: \{name.surname\}@dei.unipd.it}{}}
%\setcopyright{none}
%\settopmatter{printacmref=false, printccs=true, printfolios=true}

\copyrightyear{2020}
\acmYear{2020}
\setcopyright{acmlicensed}
\acmConference[WNS3 2020]{2020 Workshop on ns-3}{June 17--18, 2020}{Gaithersburg, MD, USA}
\acmBooktitle{2020 Workshop on ns-3 (WNS3 2020), June 17--18, 2020, Gaithersburg, MD, USA}
\acmPrice{15.00}
\acmDOI{xxxxxxxxxxxx}
\acmISBN{xxxxxxxxxxxx}

\pagestyle{empty}

\begin{abstract}

Vehicle-to-vehicle (V2V) communications have opened the way towards cooperative automated driving as a means to guarantee improved road safety and traffic efficiency.
The use of the millimeter wave (mmWave) spectrum for V2V, in particular, holds great promise since the large bandwidth available offers the possibility of realizing high-data-rate connections.
However, this potential is hindered by the significant path and penetration loss experienced at these frequencies.
It then becomes fundamental to practically evaluate the feasibility of installing mmWave-based technologies in the vehicular scenario, in view of the strict latency and throughput requirements of future automotive applications.
To do so, in this paper we present \millicar{}, the first ns-3 module for V2V mmWave networks, which features a detailed implementation of the  sidelink Physical (PHY) and Medium Access Control (MAC) layers based on the latest NR V2X specifications, the 3GPP standard for next-generation vehicular systems.
Our module is open-source and enables researchers to compare possible design options and their relative performance through an end-to-end full-stack approach, thereby stimulating further research on this topic.

\begin{picture}(0,0)(0,-360)
\put(0,0){
\put(0,0){This paper has been submitted to WNS3 2020. Copyright may be transferred without notice.}}
\end{picture}

\end{abstract}



\maketitle

\section{Introduction}\label{sec:intro}
\glsresetall

The \gls{5g} of mobile networks will support new market verticals and applications, thanks to a flexible design of the network architecture and of the protocol stack~\cite{38300}. One of the fundamental novelties of NR, i.e., the \gls{3gpp} \gls{ran} for 5G, is the communication at \gls{mmwave} frequencies. \gls{3gpp} NR, indeed, supports a carrier frequency up to 52.6 GHz in Release 15, with possible extensions to the even higher spectrum planned for Release 16 and 17. The usage of \glspl{mmwave} enables unprecedent data rates (in the order of multi gigabit per second) in the access network~\cite{akoum2012coverage}, thanks to the availability of large chunks of free spectrum (a single carrier in NR can operate on a bandwidth of up to 400 MHz).

In particular, the \gls{mmwave} bands support the most bandwidth-hungry applications in a vehicular context, where sensors (e.g., radars, LiDARs, cameras) and infotainment systems are expected to generate data at a rate of Terabytes per driving hour~\cite{choi2016millimeter}.
%The exploitation of these frequency bands is being considered also for new uses cases, besides the cellular radio access. Notably, the multi-gigabit-per-second throughput that can be achieved at \glspl{mmwave} has been seen as an opportunity for bandwidth-hungry applications in a vehicular context, where sensors (e.g., radars, lidars, cameras) and infotainment systems are expected to generate data in the order of hundreds of megabit per second~\cite{choi2016millimeter}.
Standardization bodies have already started considering the integration of \glspl{mmwave} in the specifications for next-generation vehicular networks, such as IEEE 802.11~\cite{TGbd.general} and 3GPP NR V2X~\cite{3gpp.38.885}.

The communication at \glspl{mmwave} in a vehicular scenario, however, introduces numerous challenges that need to be carefully addressed to guarantee a high quality of service for the end users and make such systems reliable and robust to the vehicle mobility~\cite{giordani2017millimeter}.
First, the severe propagation loss experienced at high frequencies prevents long-range communications. This can be partially compensated using large antenna arrays, that can focus the transmitted power in sharp beams and increase the link budget. The usage of directional transmissions, however, implies a coordination between the two communication endpoints, which have to continuously align their beams to experience a sufficient beamforming gain~\cite{kutti2016beamforming}. Moreover, \gls{mmwave} signals can be easily blocked by common obstacles such as, for example, vehicles, road signs, pedestrians. Additionally, the metal of the vehicle bodies acts as a strong reflector for \glspl{mmwave}, creating strong and bursty interference~\cite{petrov2018impact}. The interaction of the \gls{mmwave} channel with the vehicular environment also creates an extremely dynamic channel, with a fluctuating capacity and link availability, that may have a serious impact on the whole protocol stack~\cite{giordani2019lte}.

These challenges become even harder in a \gls{v2v}\footnote{While 3GPP refers to NR-V2X, i.e., the incorporation of Vehicle-to-(Roadway)Infrastructure (V2I), Vehicle-to-Network (V2N), Vehicle-to-Vehicle (V2V), Vehicle-to-Pedestrian (V2P) and Vehicle-to-Device (V2D), our effort is focused on the design of V2V protocols. From now on, when we refer to the 3GPP standard, we will use NR-V2X and when we described our module's operations we will use V2V terminology.} context, which, in principle, should be designed to work also in the absence of the fixed network infrastructure that can be exploited, instead, in a \gls{v2i} use case. The research community has started proposing solutions to improve the performance of \gls{v2v} communications at \glspl{mmwave}~\cite{perfecto2017millimeter,mavromatis2018efficient}. As of today, the lack of testbeds for \gls{mmwave} \gls{v2v} scenarios makes simulation the preferred means for the performance evaluation of novel networking designs. However, to the best of our knowledge,  an open-source, publicly available network simulator that integrates \glspl{mmwave} and \gls{v2v} scenarios is not currently available: simulation tools for \glspl{mmwave}, indeed, only support fixed infrastructure scenarios~\cite{mezzavilla2018end,PATRICIELLO2019101933}, while simulators for ad-hoc communications (e.g., \gls{d2d}) only model sub-6 GHz frequencies~\cite{rouil2017implementation}.

To fill this gap, in this paper we introduce \millicar{}, an open-source ns-3 module for \gls{v2v} \gls{mmwave} networks\footnote{Available at \url{https://github.com/signetlabdei/millicar}}. The  module introduces a characterization of  sidelink \gls{phy} and \gls{mac} layers that follow the \gls{3gpp} numerologies for NR V2X~\cite{3gpp.38.885}, and enables the study and development of beamforming, link-adaptation and medium access techniques for \gls{mmwave} \gls{v2v} in end-to-end, full-stack simulations. Additionally, the module features the \gls{3gpp} channel model introduced in~\cite{3gpp.37.885}, which has been designed for vehicular simulations in urban and highway scenarios. \millicar{} integrates the \gls{lte} \gls{sap} to connect the \gls{mac} layer to \gls{rlc} and \gls{pdcp}, and implements a new ns-3 \texttt{NetDevice} (i.e., \texttt{MmWaveVehicularNetDevice}) to take care of the integration with the TCP/IP stack of ns-3. Finally, the module incorporates (i) a helper, that can be used to easily set up simulations; (ii) unit tests, to guarantee that the module behaves as expected even when adding new features; and (iii) several examples, to simulate scenarios with a varying number of vehicles and different deployments.

\millicar{} was developed as a standalone module (e.g., with respect to the other ns-3 modules that support \glspl{mmwave}~\cite{mezzavilla2018end,PATRICIELLO2019101933}) to separate the sidelink implementation from that of scheduled cellular protocol stacks. One of the design goals of this module, indeed, is a lean implementation, and the possibility to extend it with new features without having to deal with the complexity of protocol stacks that have not been designed from the start to support a sidelink. Nonetheless, the \millicar{} module can still rely on the higher layers from the \gls{lte} module through \glspl{sap}, to run end-to-end simulations with an NR-V2X-like protocol stack. We believe that this constitutes a good tradeoff between integration with ns-3 and flexibility to develop new components.

We also report a preliminary performance evaluation, to validate the main functionalities of the module, with throughput, latency and \gls{sinr} results in different scenarios, varying system parameters such as the bandwidth, source rate and mobility of vehicles.

The remainder of this paper is organized as follows. In Sec.~\ref{sec:vehi}, we provide an overview on the research and standardization activities related to vehicular networks. The module is described in Sec.~\ref{sec:implementation}, where we comment on the channel model and the sidelink physical and \gls{mac} layers. We discuss the structure of tests and helpers in Sec.~\ref{sec:helpers}, while the example scenarios are introduced in Sec.~\ref{sec:scenarios}. Finally, in Sec.~\ref{sec:conclusions} we conclude the paper and provide suggestions for future extensions.

\section{Next-Generation Vehicular Networks}
\label{sec:vehi}

The recent evolution of hardware, software, and communication technologies in the automotive sector has paved the way towards \glspl{c-its} as a means to support traffic safety, traffic efficiency and infotainment services.
More specifically, \glspl{c-its} promise to reduce the number of road accidents and carbon emission by more than 90\% and 60\% respectively, as well as to
improve  drivers' productivity during commutes~\cite{clements2017economic}. Overall, the market potential of \glspl{c-its} is estimated in 1.3 trillion USD annually in the U.S. alone, thereby stimulating further research in the field of next-generation vehicular networks.

\begin{figure*}[t!]
	\setlength\belowcaptionskip{-0.3cm}
	\centering
	\includegraphics[width=.99\textwidth]{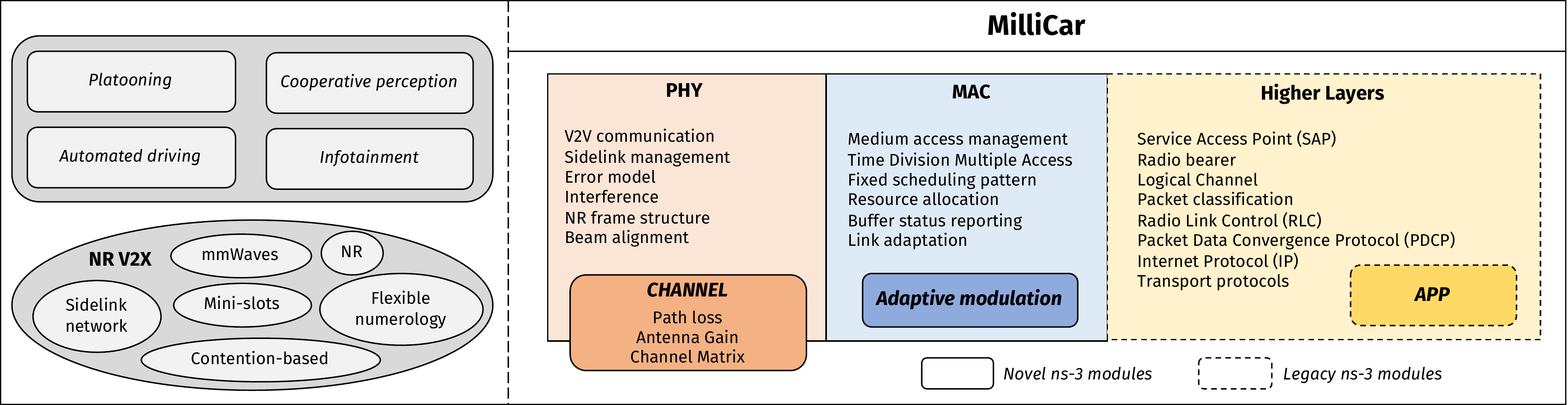}
	\caption{General overview of the \millicar{} module, with the features implemented at each layer of the stack.}
	\label{fig:mod}
\end{figure*}

When fully deployed, \glspl{c-its} will address the demands and business trends of the 2030 society, and support the new use cases highlighted in Fig.~\ref{fig:mod}. They include (but are not limited to):
\begin{itemize}
	\item \emph{platooning}, where vehicles can travel in close proximity to one another at very high speeds;
	\item \emph{cooperative perception}, where vehicles broadcast sensors' observations to increase the perception range of their on-board~instrumentation;
	\item \emph{semi- or fully-automated driving}, where vehicles can sense the surrounding environment and move with minimal or no human input;
	\item \emph{infotainment}, i.e.,  a set of services that deliver a combination of information and entertainment.
\end{itemize}
These applications have very strict demands in terms of data rate (in the order of Terabytes per driving hour, according to some estimates), latency (from approximately 3 ms up to 100 ms depending on the degrees of automation) and reliability (up to 99.999\% for the most safety-critical services)~\cite{3GPP_22186}. In order for these heterogeneous requirements to be satisfied, \glspl{c-its} will enable wireless communications to and from roadside infrastructures and among vehicles, a concept that is referred to as \gls{v2x} connectivity.
While current  \gls{v2x} standard technologies, i.e.,  IEEE 802.11p and 3GPP Cellular-V2X (C-V2X), may lack the level of reliability and availability requested by future vehicular applications, the IEEE and 3GPP are promoting standardization efforts, i.e.,  802.11bd~\cite{TGbd.general} and NR V2X~\cite{3gpp.38.885} respectively, to overcome current technology limitations~\cite{zugno2019towards}, as described in the following paragraphs.

\paragraph*{IEEE 802.11bd} The 802.11bd standard will enhance IEEE 802.11p targeting future V2X application requirements.  Although technical details have not yet been thoroughly discussed, the 802.11bd evolution is positioned to support reduced latency and twice the throughput and the communication range of 802.11p  through (i) new transmission mechanisms,  (ii) dual carrier modulation with midambles to provide a better channel estimation in fast-fading channels~\cite{TGbd.numerologies}, and (iii) re-designed PHY and MAC features (including a flexible sub-carrier spacing, with up to 40 MHz channel bandwidth).

\paragraph*{3GPP NR V2X} The NR V2X specifications will amend 3GPP C-V2X as part of the ongoing 5G effort. New developments will include (i) more sophisticated channel models, (ii) sidelink and network architecture improvements, (iii)  support of mini-slot scheduling for latency-critical services, (iv) a flexible numerology, along the lines of the 3GPP Rel. 15 specifications, and (v) new resource allocation schemes where sidelink resources are either  scheduled by the base station (mode 1) or autonomously by the vehicles (mode 2).\\

In particular, both IEEE 802.11bd and 3GPP NR V2X standards will support operations at \gls{mmwave} frequencies, i.e., in the 57-71~GHz and 24.25-52.6~GHz ranges, respectively.
As discussed in Sec.~\ref{sec:intro}, the large spectrum availability at these frequencies promotes high-capacity low-latency communication, however working at \gls{mmwave} raises several challenges from a communication point of view.
For this reason, it is important to demonstrate the practical feasibility of designing protocols to support vehicular operations at such high frequencies.

While testbed validation is impractical due to the prohibitive costs of the \gls{mmwave} hardware and due to the difficulties associated to the design and implementation of a representative and not-too-simple application scenario, end-to-end simulation is the preferred solution for performance evaluation.
Currently, \gls{v2i} communications  can be simulated using the ns-3 \gls{mmwave} module designed by NYU and the University of Padova (UNIPD) in 2015~\cite{mezzavilla2018end} or 5G Lena~\cite{PATRICIELLO2019101933}, which implement a complete 3GPP-like protocol stack, but focus on cellular-like, infrastructure-based deployments.
For the \gls{v2v} scenario, instead, there are currently no open-source simulators capable of modeling the \gls{mmwave} channel as well as a full TCP/IP protocol stack and the mobility of vehicles.
Specifically, traditional vehicular simulator, e.g., Veins~\cite{sommer2011bidirectionally} or V2X Simulation Runtime Infrastructure (VSimRTI)~\cite{schunemann2008novel}, support development, training, and validation of autonomous urban driving systems, but do not support mmWave communications.
% Also, the NYU/UNPD ns-3 mmWave module is based on the NR cellular standard, which  is however infrastructure-based and does not characterize infrastructure-less (ad hoc) scenarios.

In this paper, we fill this gap by providing the first open-source ns-3 module for V2V networks operating at \glspl{mmwave}, so as to enable realistic full-stack simulations in the V2V environment, as described in the next sections.

\section{A Module for NR V2X}
\label{sec:implementation}
% general structure of the module
% describe the dependencies (mmWave, LTE, spectrum - hopefully :))

\begin{figure*}[t!]
	\centering
	\includegraphics[width=0.9\textwidth]{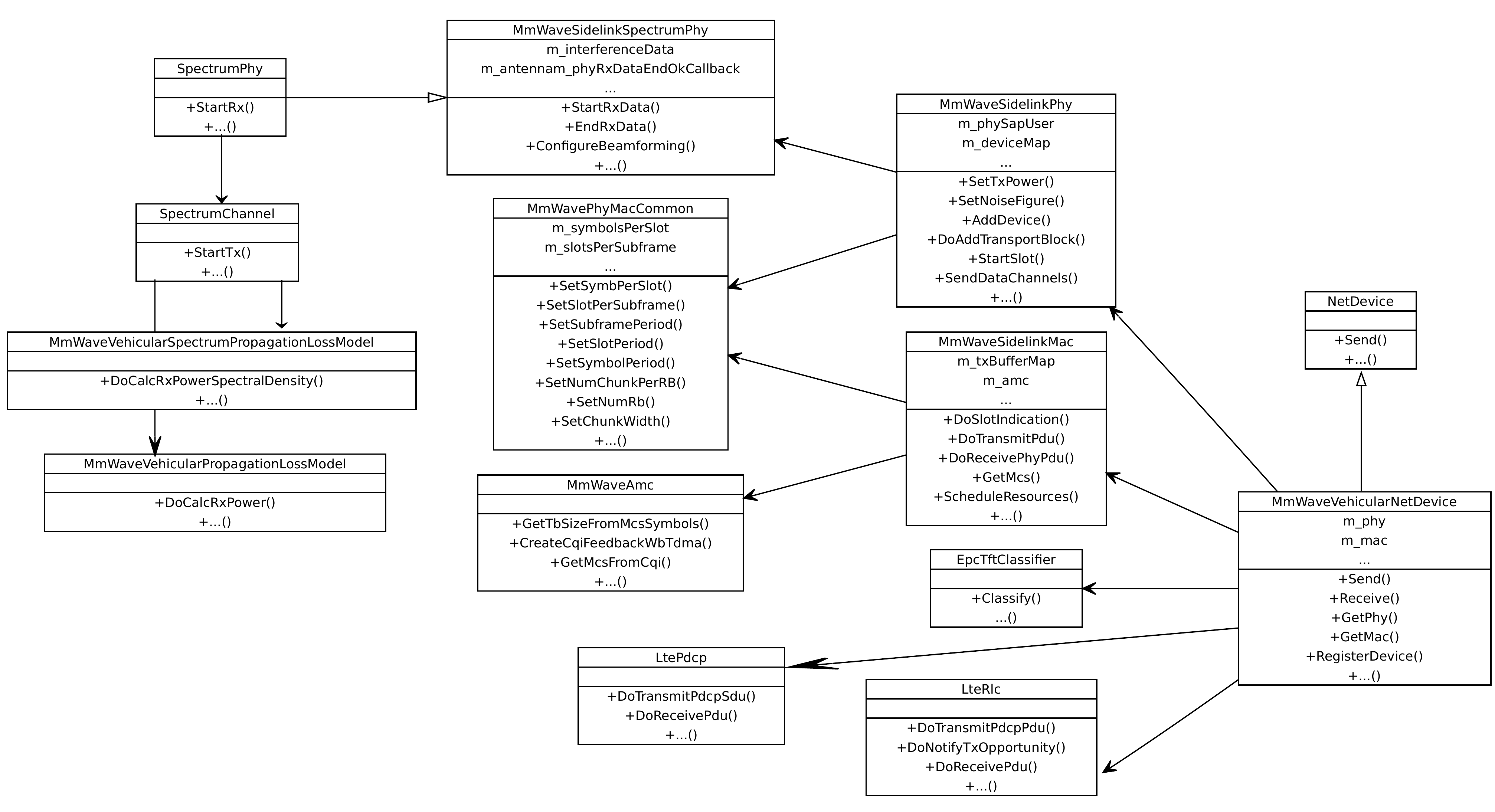}
	\caption{Simplified \gls{uml} diagram of the \millicar{} classes.}
	\label{fig:uml}
\end{figure*}

In this section we describe the main characteristics of the \millicar{} module for V2V networking. The general features of each component of the module are depicted in Fig.~\ref{fig:mod}, while Fig.~\ref{fig:uml} provides a simplified \gls{uml} diagram.

In Sec.~\ref{sec:channel} we present the channel model proposed by the 3GPP to characterize the V2V propagation at \glspl{mmwave}. Then, in Secs.~\ref{sub:phy} and \ref{sub:mac} we present the key elements of the \millicar{} \gls{phy} and \gls{mac} layers, respectively. They provide functionalities for packet transmission and reception over an NR-V2X-compliant frame structure, and a proper scheduling of the radio resources.
As mentioned in Sec.~\ref{sec:vehi}, the study item \cite{3gpp.38.885} considers both in-coverage (mode 1) and out-of-coverage (mode 2) options for resource allocation.
The protocol stack of \millicar{} natively supports mode 2, the more likely to be implemented in an early deployment of NR V2X, given that mode 1 would require an update of base stations following standard specifications~\cite{zugno2019towards,3gpp.38.885}. Also, mode 2 is of particular interest for researchers since it poses several challenges that remain to be addressed.
%, thus motivating the need for new tools and methodologies. Therefore, we designed the protocol stack of \millicar{} to natively support mode 2.

Finally, in Sec.~\ref{sub:higher-layers} we describe how the module integrates with the higher layers of the ns-3 protocol stack. In particular, the module currently relies on the implementation of the \gls{rlc} and \gls{pdcp} layers and of the traffic flow filters of the \texttt{lte} module.

Notice that \millicar{} is integrated with ns-3, and reuses a number of data structures and classes (for example - to hold the configuration parameters of the frame structure) from the NYU/UNIPD \texttt{mmwave} module for cellular communications. Finally, we plan to complete the integration with the new features of the \texttt{spectrum} module, described in~\cite{zugno2020implementation}, which would allow us to separate the code base of the channel model from that of the protocol stack.

\subsection{Vehicular Channel Model Implementation}
\label{sec:channel}

The accurate characterization of the channel behavior is paramount to obtain reliable simulation results. Therefore,  \millicar{} implements the propagation and fading models that the \gls{3gpp} suggests for \gls{v2v} communications at \glspl{mmwave}~\cite{giordani2019pathloss,3gpp.37.885}, by means of the APIs provided by the ns-3 \texttt{Spectrum} module. In particular, devices communicating through the same wireless channel are attached to a single instance of \texttt{SpectrumChannel} which accounts for the modeling of the propagation phenomena using the interfaces \texttt{PropagationLossModel} and \texttt{SpectrumPropagationLossModel}.

The  pathloss has been implemented in \texttt{MmWaveVehicularPropagationLossModel}, which extends the \texttt{PropagationLossModel} interface and adopts the structure described in \cite{zhang2017channel3gpp}. Following the standard's guidelines \cite{3gpp.37.885}, we defined two scenarios, \texttt{V2V-Highway} and \texttt{V2V-Urban}, where a pair of vehicles can be in one of the following states:
\begin{compactitem}
	\item \gls{los}, if the vehicles are in the same street and the path is free from obstacles;
	\item \gls{nlos}, if the path is blocked by static objects, e.g., buildings;
	\item \gls{nlosv}, if the vehicles are in the same street but the path is blocked by other vehicles.
\end{compactitem}

This state can be fixed during the whole simulation by setting the \texttt{ChannelCondition} attribute. By default, if not specified, the \gls{los} state is updated when the overall channel characterization is updated. Notice that the model in~\cite{3gpp.37.885} does not provide a closed-form probabilistic expression for the transition from \gls{los} or \gls{nlosv} to \gls{nlos}. Indeed, the model first checks if the vehicles are in different streets, and, if this is the case, it deterministically sets the channel condition to NLOS. Otherwise, for communications between vehicles that are in the same street, the \gls{los} (or \gls{nlosv}) condition is determined following the state-transition probability described in  \cite[Sec. 6.1]{3gpp.37.885}.

% This state can be fixed during the whole simulation by setting the \texttt{ChannelCondition} attribute. By default, if not specified, the \gls{los} state is updated when the overall channel characterization is updated, and is determined following the state-transition probability described in  \cite[Sec. 6.1]{3gpp.37.885}.
For the \texttt{V2V-Highway} scenario, the \gls{los} probability  is given by
\begin{equation}
	P_{\rm LOS} =
	\begin{cases}
	\min \left \{ 1, a\cdot d^2 + b\cdot d + c \right \} & d \leq 475 \text{ m}    \\
	\max \left \{0, 0.54-0.001\cdot (d-475) \right \} & d > 475 \text{ m},   \\
	\end{cases}
\end{equation}
where $a = 2.1013 \cdot 10^{-6}$, $b = -0.002$, $c = 1.0193$, and $d$ is the distance between the vehicles, measured in meters.
For the \texttt{V2V-Urban} scenario, instead, the  \gls{los} probability is evaluated as
\begin{equation}
	P_{\rm LOS} = \min \left\{ 1, 1.05 \cdot  e^{-0.0114\cdot d} \right\}.
\end{equation}
In both cases, $P_{\rm NLOSv} = 1 - P_{\rm LOS}$.

% The model in~\cite{3gpp.37.885} does not provide a closed-form probabilistic expression for the transition to \gls{nlos} transition as it is deterministically set for vehicles which are not in the same street.
To also account for a probabilistic characterization of the \gls{nlos} state, the \millicar{} module defines two additional scenarios, namely \texttt{Extended-V2V-Highway} and \texttt{Extended-V2V-Urban}, that implement the state probabilities detailed in \cite{boban2016evolution}.
For the \texttt{Extended-V2V-Highway} scenario,  the \gls{los} and \gls{nlos} probabilities are given by
\begin{equation}
	P_{\rm LOS} = \min\Big\{1,\max\{0,2.7 \cdot 10^{-6}\cdot d^2-0.0025\cdot d+1\}\Big\},
\end{equation}
\begin{equation}
	P_{\rm NLOS} = \min\Big\{1,\max\{0,-3.7 \cdot 10^{-7}\cdot d^2+0.00061\cdot d+0.015\}\Big\}.
\end{equation}
\begin{equation}
	P_{\rm NLOSv} = 1 - P_{\rm LOS} - P_{\rm NLOS}.
\end{equation}

For the \texttt{Extended-V2V-Urban} scenario, instead, the \gls{los} and \gls{nlos} probabilities are equal to
\begin{equation}
P_{\rm LOS} = \min\{1,\;\max\{0, 0.8372\cdot e^{-0.0114\cdot d}\}\},
\end{equation}
\begin{equation}
P_{\rm NLOSv} = \min\left\{1,\;\max\left\{0, \frac{1}{0.0312\cdot d}\cdot e^{\frac{(-(\ln(d) - 5.0063))^2}{2.4544}}\right\}\right\},
\end{equation}
\begin{equation}
P_{\rm NLOS} = 1 - P_{\rm LOS} - P_{\rm NLOSv}.
\end{equation}
All these terms represent a vehicular traffic condition of medium density (i.e., 120 vehicles/km$^2$ in the urban scenario, and 1500 vehicles/h/dir in the highway scenario).

For each scenario, the link pathloss is implemented accurately following the 3GPP specifications~\cite{3gpp.37.885}. For the highway scenarios (i.e., \texttt{V2V-Highway} and \texttt{Extended-V2V-Highway}), the \gls{los} pathloss~is
\begin{equation}
		PL_{\rm LOS} = 32.4 + 20*\log_{10}{d} + 20*\log_{10}{f_c} \; [dB],
\end{equation}
where the distance $d$ is in meters and the center frequency $f_c$ is in GHz. For the urban scenarios (i.e., \texttt{V2V-Urban} and \texttt{Extended-V2V-Urban}), instead, the LOS pathloss is given by
\begin{equation}
		PL_{\rm LOS} = 38.77 + 16.7*\log_{10}{d} + 18.2*\log_{10}{f_c} \; [dB].
\end{equation}
In case of NLOS, the model does not distinguish between urban and highway propagation, so that the pathloss equation~becomes
\begin{equation}
		PL_{\rm NLOS} = 36.85 + 30*\log_{10}{d} + 18.9*\log_{10}{f_c} \; [dB].
\end{equation}
When the link is obstructed by a vehicle (i.e., in \gls{nlosv} state), an incremental shadowing term, modeled as a log-normal random variable, is added to the $PL_{\rm LOS}$ equation. This term depends on the blocker height, which is randomly selected among three possible vehicle types (to be configured in the simulation), as detailed in~\cite{3gpp.37.885}.
%\textr{credo che le equazioni con i termini aggiuntivi di NLOSv siano inutili, visto che poi i risultati che mostriamo sono solo con LOS fissata.. ma se secondo voi vanno aggiunti, eseguo}

The operations associated to fast fading and beamforming gain computation have been implemented in the class \texttt{MmWaveVehicularSpectrumPropagationLossModel} which extends the interface of \texttt{SpectrumPropagationLossModel} to guarantee consistency among different channel models provided by ns-3. The procedure to generate the channel is based on the  specifications in~\cite{3gpp.38.901}, while the dual mobility (to compute the Doppler effect) and the parameters of the equations that define the channel are modeled as described in~\cite{3gpp.37.885}. Also, to take into account the non-isotropic behavior of real antennas, we extended the class \texttt{AntennaArrayModel} of the \texttt{mmwave} module, including the radiation model specified in~\cite{3gpp.37.885}. This class handles also the computation of the beamforming vectors using a \gls{dft} based approach~\cite{yang2010dftbeam}. %As a consequence, the power spectral density at the receiver is computed inside the implementation of the \texttt{DoCalcRx\-Power\-SpectralDensity} method.

\vspace{-.2cm}
\subsection{Physical Layer}
\vspace{-.1cm}
\label{sub:phy}
% composed by two classe{}s
The \millicar{} physical layer is composed of two classes, namely, \texttt{MmWave\-SidelinkSpectrumPhy} and \texttt{MmWaveSidelinkPhy}.
% mmwavasidelinkspectrumphy takes care of generating the signal (create the psd)
% interface between the channel and the mmwave vehicular net device
% handle the signal reception
% discard signals that are not meant for this dev
% interacts with mmwave interference to compute the sinr
% apply the error model
% generate the sinr reports
\texttt{MmWave\-Sidelin\-kSpectrumPhy} extends the abstract class \texttt{SpectrumPhy} and acts as an interface between the \texttt{MmWaveVehicularNetDevice} and the \texttt{SpectrumChannel}. In fact, it handles the transmission and reception operations through the methods \texttt{StartTxDataFrames} and \texttt{StartRx}.
The method \texttt{StartTxDataFrames} generates the signal to be transmitted over the channel, represented by the structure \texttt{MmWaveSidelinkSpectrumSignalParameters}. Then, it forwards it to the \texttt{SpectrumChannel} instance by calling the method \texttt{SpectrumChannel::StartTx}.
% Instead, when a signal is received from the \texttt{SpectrumChannel}, the method \texttt{StartRx} checks if can be decoded based on or not and, if so, decides about the outcome by applying the error model.
Conversely, when a signal is received from the \texttt{SpectrumChannel}, the method \texttt{StartRx} checks whether or not it can be decoded  by applying an error model and, if so, forwards it to the upper layer.
The error model that is currently supported by our module is based on the one described in \cite{mezzavilla2018end}, which derives the error probability taking as input the received \gls{sinr} and the \gls{mcs} used to encode the signal.
To compute the  \gls{sinr}, \texttt{MmWaveSidelinkSpectrumPhy} relies on the classes \texttt{mmWaveInterference} and \texttt{mmWaveChunkProcessor}, of  the \texttt{mmwave} module~\cite{mezzavilla2018end}.
Moreover, \texttt{MmWaveSidelinkSpectrumPhy} takes care of the periodic generation of the \gls{csi} reported to the upper layers.

%mmwavesidelinkspectrumphy is in charge of maintaining the synchronization marking frame, subframe, slots

The class \texttt{MmWaveSidelinkPhy} is in charge of maintaining the system synchronization, and of managing the physical channel used for the transmission and reception of the transport blocks (our module currently supports the modeling of \gls{pssch} only).
The frame structure used by \millicar{} is compliant with NR specifications, i.e., a frame of $10$ ms is divided in $10$ subframes, each containing a variable number of slots. Each slot is composed of $14$ \gls{ofdm} symbols, whose duration depends on the selected numerology configuration~\cite{38211}.
Following the proposal in~\cite{3gpp.38.885}, our module currently supports NR numerologies 2 and 3, i.e., with 4 and 8 slots per subframe, respectively, corresponding to a \gls{scs} of 60 kHz or 120 kHz.
A transmission buffer is used to store the transport blocks to be sent during the first available slot, together with information regarding the \gls{mcs} to use and the allocated \gls{ofdm} symbols. %and can be filled through the \texttt{AddTransport} method.
%a simulator-specific class (PacketBurst) is used to aggregate MAC SDUs in order to achieve the simulator’s equivalent of a TB, without the corresponding implementation complexity. The multiplexing of different logical channels to and from the RLC layer is performed using a dedicated packet tag (LteRadioBearerTag), which performs a functionality which is partially equivalent to that of the MAC headers specified by 3GPP
\begin{figure}[t!]
  \setlength{\belowcaptionskip}{-0.33cm}
	\centering
	\includegraphics[trim={0cm, 2cm, 0cm, 3cm},width=.49\textwidth]{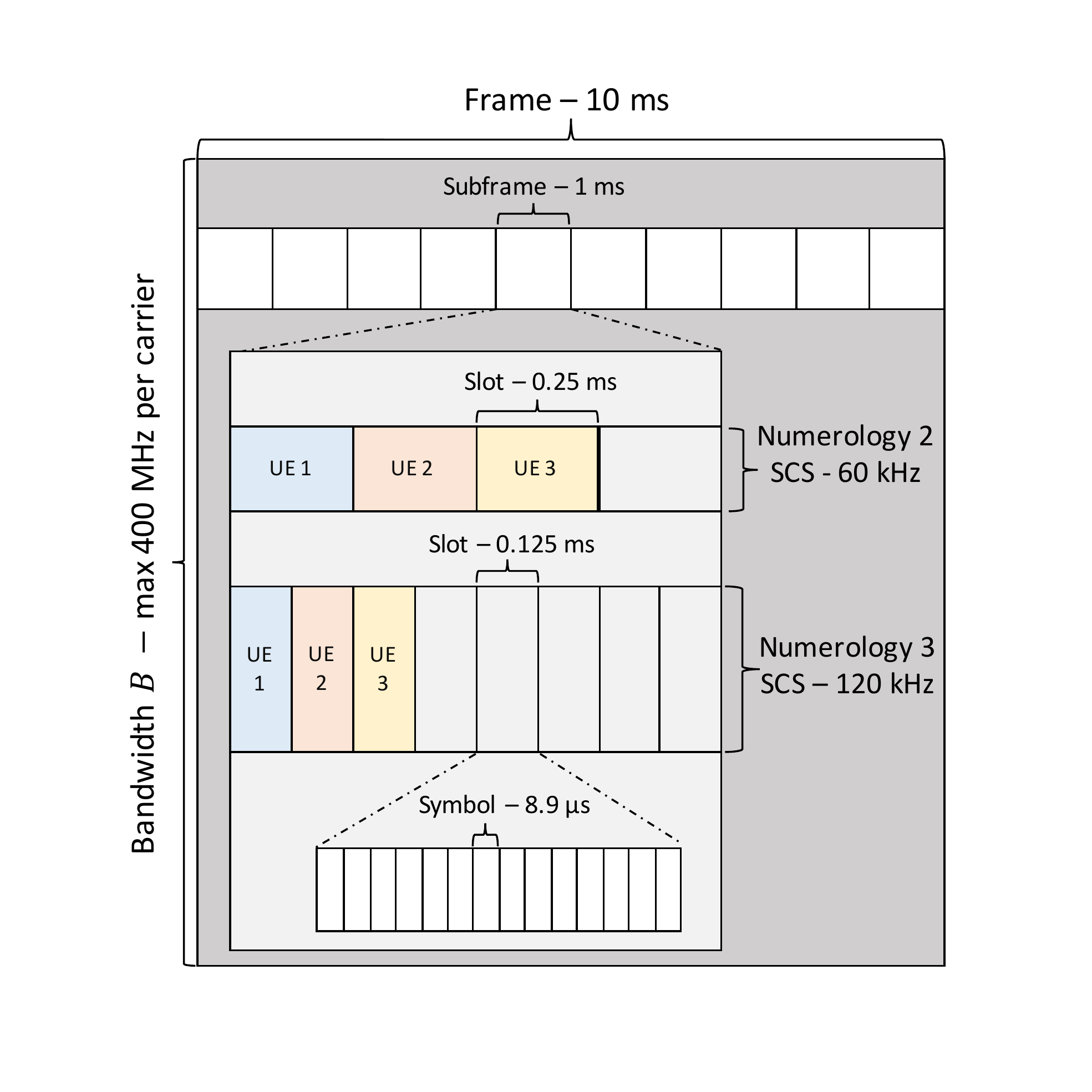}
	\caption{Frame structure configurations supported by \millicar{}. The different colors represent a possible allocation pattern.}
	\label{fig:frame}
  % \vspace{-0.45cm}
\end{figure}

The method \texttt{StartSlot} marks the beginning of each slot and takes care of transmitting the transport blocks stored in the buffer, by scheduling multiple calls to \texttt{MmWaveSidelinkSpectrumPhy::Start\-TxDataFrames}.
% The beginning of each slot is marked by the method \texttt{StartSlot} , which takes care of transmitting the data in the buffer by scheduling multiple calls to \texttt{MmWaveSidelinkSpectrumPhy::StartTxDataFrames}, one for each transport block.
Moreover, \texttt{MmWaveSide\-linkPhy} takes care of forwarding the received transport blocks to the upper layer and managing the beamforming operations to properly point the beam towards the other end device (at this stage, perfect beam alignment is assumed, and further refinements are left for future work).

\subsection{MAC Layer}
\label{sub:mac}
The \gls{mac} layer functionalities are implemented in the  \texttt{MmWaveSide\-linkMac} class, which includes:
% the management of the medium access and the multiplexing of multiple logical channels onto the physical channel by proper scheduling of the available resources. % link adaptation
 (i) the management of the medium access, (ii) the scheduling of the available resources, (iii) the support of transmission and reception over multiple logical channels, and (iv) the link adaptation.

% \subsubsection{Management of the medium access}

\texttt{MmWaveSidelinkMac} supports a \gls{tdma}-based access scheme, where different vehicles transmit in different slots, as generally assumed  for directional \gls{mmwave} operations~\cite{mezzavilla2018end}.
Similarly to mode 2c defined in \cite{3gpp.38.885},
%each vehicle is pre-configured with a fixed scheduling pattern, which defines the slots that can be used for transmission.
the \gls{mac} layer is pre-configured with a fixed scheduling pattern, which determines how the slots are assigned to the vehicles on a per-subframe basis.
By default, each vehicle can use a single slot per subframe, but this pattern can be customized using the \texttt{SetSfAllocationInfo} method.

% \subsubsection{Scheduling of the resources}
At the beginning of each slot, the \gls{mac} layer retrieves the scheduling pattern and executes \texttt{DoSlot\-Indication} to  decide whether to perform the transmit or receive operation.
% In case the slot is intended for transmission, the available resources are equally divided among the active data streams by means of the method \texttt{ScheduleResources} and notified to the upper layer.
% The scheduling decision is notified to the the upper layer which forwards the the \glspl{sdu} to be transmitted
In case the slot is intended for transmission, the method \texttt{ScheduleResources} divides the available resources among the active logical channels\footnote{A logical channel represents an end-to-end connection at the \gls{mac} and physical layers.} and notifies the scheduling decision to the upper layers.
Then, it builds the transport block using the \glspl{sdu} received from the higher layers, which is then forwarded to the \gls{phy} layer by calling the method \texttt{AddTransportBlock}.
To avoid the allocation of unnecessary resources, the buffers at the upper layers are monitored through a periodic buffer status reporting procedure. Such reports are used to decide the amount of resources to be reserved for each logical channel.
Conversely, if the slot is dedicated to another device, the \gls{phy} layer is informed about a possible incoming reception by the \gls{mac}, which then performs de-multiplexing operations to map the received packets onto the proper logical channels.

% \subsubsection{Link Adaptation}
Moreover, automatic link adaptation functionalities are provided based on \gls{csi} reports received from the \gls{phy}.
%The reports are cached in a map?
This mechanism is handled by the \texttt{MmWaveAmc} class, which uses the last received \gls{csi} report to determine the optimal modulation and coding scheme to be used for the transmission.

% tdma
% fixed scheduling pattern
% scheduling of the resources between the logical channels
% holds the status o of the rlc buffers
% forward the mac pdus to the phy buffer
% fixed mcs or adaptive modulation and coding (the channel quality is estimated from the received messages)

\subsection{Integration with the higher layers}
\label{sub:higher-layers}
\millicar{} also provides full integration with the higher layers of the protocol stack ensuring, by means of \gls{sap}, high flexibility for future improvements of the stack design. We attach to each \texttt{MmWaveSide\-linkMac} instance occurrences of \texttt{LteRlc}, which is then linked to an \texttt{LtePdcp} object. Closing the gap, a specific class implementing the \gls{sap} is used to connect the \gls{pdcp} object to our ad hoc \texttt{MmWaveVehicularNetDevice}, envisioning in this way a full bottom-up and top-down integration. The instances for these layers for each end-to-end connection are managed inside the \texttt{MmWaveVehicularNetDevice} class, which extends \texttt{NetDevice} and implements all the virtual classes commonly used to set up the communication to and from the TCP/IP stack.

In order for two nodes to communicate, a radio bearer must be set up. Once a \texttt{MmWaveVehicularNetDevice} is associated to each node, the method \texttt{MmWaveVehicularNetDevice::Activate\-Bearer} is executed on both communication endpoints. This function accepts as input an integer number representing the \texttt{bearerID}, the \gls{rnti} of the destination (an integer number that differentiates distinct nodes in the network) and the IP address of the pairing node. In particular, each \texttt{bearerID} must unequivocally identify a radio bearer, and cannot be shared among different pairs of devices. The consistency of this assignment (along with that of the \gls{rnti}) among different nodes is guaranteed in the helper's configuration method, which will be described in Sec.~\ref{sec:helpers}. At the \gls{mac} layer, a bearer is mapped to a logical channel identifier, as defined by the 3GPP standard. However, at this stage of development, logical channels and radio bearers have a one-to-one correspondence.

The operations carried out by the \texttt{ActivateBearer} method are:
\begin{itemize}
	\item the creation of a rule to classify packets generated from different sources, using \texttt{EpcTft::PacketFilter};
	\item the instantiation of an \texttt{LteRlc} object, which can be identified by the \gls{rnti} of the destination node and the logical channel identifier. The \gls{rlc} object is then linked to the \gls{mac} layer instance associated to the node;
	\item the creation of an \texttt{LtePdcp} object, which has to be connected to the \texttt{MmWaveVehicularNetDevice} and the \gls{rlc} object created in the previous step.
\end{itemize}
After these steps, the \gls{rlc} and \gls{pdcp} objects are stored in a dedicated structure, i.e., \texttt{SidelinkRadioBearerInfo}, which is then identified with the univocal \texttt{bearerID} and saved in the \texttt{m\_bearerToInfoMap} variable. Currently, the version of the \gls{rlc} supported by this module is \texttt{LteRlcUm}, which provides segmentation and concatenation but no retransmissions.

Once \texttt{MmWaveVehicularNetDevice::Send} receives a packet from the IP layer, it accesses the \texttt{m\_tftClassifier} variable to retrieve the \texttt{bearerID} that associates the \gls{rnti} and the logical channel identifier to the packet, and stores them in the \texttt{LtePdcpSapProvider\-::TransmitPdcpSduParameters} struct. This is then forwarded to the \gls{pdcp}. Conversely, on the reception phase, a packet is simply sent from \gls{pdcp} to the \texttt{NetDevice}, and from the \texttt{NetDevice} to the upper layers.

\section{Helpers and Test Framework}
\label{sec:helpers}

The mmWave vehicular module is also equipped with \emph{helpers} (Sec.~\ref{sub:helper}) to allow the users to easily set up the simulation, and \emph{unit tests} (Sec.~\ref{sub:test}) to  check basic functionalities of the module and facilitate future class developments.

\subsection{Helpers}
\label{sub:helper}
The main helper is the \texttt{MmWaveVehicularHelper} which (i) creates and configures the objects for the channel computation; (ii) computes the parameters for the frame structure, according to the selected 3GPP numerology; (iii) installs the networking stack on the vehicles; and (iv) connects groups of vehicles that will communicate together. The first operation is performed during initialization, and relies on three \texttt{StringValue} attributes that configure the propagation loss model (\texttt{PropagationLossModel}), the fading model (\texttt{SpectrumPropagationLossModel}), and the propagation delay model (\texttt{PropagationDelayModel}).
A typical configuration would include the propagation and fading classes described in Sec.~\ref{sec:channel}, without a delay model, as this is included in the spectrum model. However, the user can change and select different options (e.g., a simple Friis propagation loss) and combine also a delay model. The \texttt{Numerology} attribute, which is linked to the \texttt{SetNumerology} method, accepts 2 or 3 as integer value, to select among the two different numerologies currently foreseen for NR V2X.

The method \texttt{InstallMmWaveVehicularNetDevices} accepts a container of \texttt{Node} objects, and returns the \texttt{NetDeviceContainer} with the \texttt{MmWaveVehicularNetDevice} objects. Additionally, for each vehicle, this method sets up the instances of the \gls{phy} and \gls{mac} layers, configures the antenna at the vehicle and connects it to the channel.

\begin{figure*}
  \setlength{\belowcaptionskip}{-0.33cm}
\begin{subfigure}{0.48\textwidth}
	\centering
	\setlength\fwidth{\columnwidth}
	\setlength\fheight{.5\columnwidth}
	% This file was created by matplotlib2tikz v0.7.4.
\begin{tikzpicture}

\definecolor{color0}{rgb}{0.12156862745098,0.466666666666667,0.705882352941177}
\definecolor{color1}{rgb}{1,0.498039215686275,0.0549019607843137}
\definecolor{color2}{rgb}{0.172549019607843,0.627450980392157,0.172549019607843}
\definecolor{color3}{rgb}{0.83921568627451,0.152941176470588,0.156862745098039}
\definecolor{color4}{rgb}{0.580392156862745,0.403921568627451,0.741176470588235}
\definecolor{color5}{rgb}{0.549019607843137,0.337254901960784,0.294117647058824}

\pgfplotsset{every tick label/.append style={font=\scriptsize}}

\begin{axis}[
width=0.951\fwidth,
height=\fheight,
at={(0\fwidth,0\fheight)},
legend cell align={left},
legend style={at={(0.03,0.97)}, anchor=north west, draw=white!80.0!black},
tick align=outside,
tick pos=left,
x grid style={white!69.01960784313725!black},
xlabel={Source Rate [Mbps]},
xmajorgrids,
xmin=0, xmax=840,
xtick style={color=black},
y grid style={white!69.01960784313725!black},
ylabel={Average Throughput [Mbps]},
ymajorgrids,
ymin=0, ymax=810,
ytick style={color=black},
ylabel style={font=\footnotesize\color{white!15!black}},
xlabel style={font=\footnotesize\color{white!15!black}},
]
\addplot [semithick, mark=*, color0]
table {%
819.2 177.94185696005
409.6 175.55881924992
182.044444444444 180.035604478854
81.92 81.0794904997719
8.192 8.15882699689761
};
% \addlegendentry{BW=100.0MHz - d=50.0 m}

\addplot [semithick, mark=*, mark options={solid,}, color0, dashed]
table {%
819.2 801.26171075925
409.6 402.342170605654
182.044444444444 176.842253848405
81.92 79.5909057497589
8.192 8.00263014367765
};
% \addlegendentry{BW=400.0MHz - d=50.0 m}

\addplot [semithick, mark=o, color1]
table {%
819.2 160.233338911002
409.6 151.145449237773
182.044444444444 164.672217546811
81.92 73.5153175641287
8.192 7.51598209995031
};
% \addlegendentry{BW=100.0MHz - d=150.0 m}

\addplot [semithick, mark=o, mark options={solid,}, color1, dashed]
table {%
819.2 678.297224095362
409.6 356.469789166708
182.044444444444 158.450812977378
81.92 74.1843105755337
8.192 7.20447475599894
};
% \addlegendentry{BW=400.0MHz - d=150.0 m}

\addplot [semithick, mark=diamond, color4]
table {%
819.2 125.470184576151
409.6 137.590107285196
182.044444444444 122.272243536847
81.92 67.0475263569025
8.192 6.82220237948273
};
% \addlegendentry{BW=100.0MHz - d=250.0 m}

\addplot [semithick, mark=diamond, mark options={solid,}, color4, dashed]
table {%
819.2 384.745987658082
409.6 214.090472244066
182.044444444444 110.973324635368
81.92 49.2433952244256
8.192 5.4842876186374
};
% \addlegendentry{BW=400.0MHz - d=250.0 m}

\coordinate (pt) at (axis cs:50,60);

\end{axis}

\node[yshift=0cm,pin={[pin edge={ultra thin}]85:{%
    \begin{tikzpicture}[trim axis left,trim axis right]
    \pgfplotsset{every tick label/.append style={font=\tiny}}
    \begin{axis}[
scale only axis,
width=0.26\fwidth,
height=.24\fheight,
legend cell align={left},
legend style={at={(0.03,0.97)}, anchor=north west, draw=white!80.0!black},
tick align=inside,
tick pos=left,
axis x line*=bottom,
axis y line*=left,
x grid style={white!69.01960784313725!black},
xmajorgrids,
xmin=0, xmax=90,
xtick style={color=black},
y grid style={white!69.01960784313725!black},
ymajorgrids,
ymin=0, ymax=90,
ytick style={color=black},
axis background/.style={fill=white},
]

\addplot [semithick, mark=*, color0]
table [row sep=crcr] {%
819.2 177.94185696005\\
409.6 175.55881924992\\
182.044444444444 180.035604478854\\
81.92 81.0794904997719\\
8.192 8.15882699689761\\
};
% \addlegendentry{BW=100.0MHz - d=50.0 m}

\addplot [semithick, mark=*, mark options={solid,}, color0, dashed]
table [row sep=crcr] {%
819.2 801.26171075925\\
409.6 402.342170605654\\
182.044444444444 176.842253848405\\
81.92 79.5909057497589\\
8.192 8.00263014367765\\
};
% % \addlegendentry{BW=400.0MHz - d=50.0 m}

\addplot [semithick, mark=o, color1]
table [row sep=crcr] {%
819.2 160.233338911002\\
409.6 151.145449237773\\
182.044444444444 164.672217546811\\
81.92 73.5153175641287\\
8.192 7.51598209995031\\
};
% % \addlegendentry{BW=100.0MHz - d=150.0 m}

\addplot [semithick, mark=o, mark options={solid,}, color1, dashed]
table [row sep=crcr] {%
819.2 678.297224095362\\
409.6 356.469789166708\\
182.044444444444 158.450812977378\\
81.92 74.1843105755337\\
8.192 7.20447475599894\\
};
% % \addlegendentry{BW=400.0MHz - d=150.0 m}

\addplot [semithick, mark=diamond, color4]
table [row sep=crcr] {%
819.2 125.470184576151\\
409.6 137.590107285196\\
182.044444444444 122.272243536847\\
81.92 67.0475263569025\\
8.192 6.82220237948273\\
};
% % \addlegendentry{BW=100.0MHz - d=250.0 m}

\addplot [semithick, mark=diamond, mark options={solid,}, color4, dashed]
table [row sep=crcr] {%
819.2 384.745987658082\\
409.6 214.090472244066\\
182.044444444444 110.973324635368\\
81.92 49.2433952244256\\
8.192 5.4842876186374\\
};
% % \addlegendentry{BW=400.0MHz - d=250.0 m}

    \end{axis}
    \end{tikzpicture}%
}}] at (pt) {};

\begin{axis}[
width=0.951\fwidth,
height=\fheight,
at={(0\fwidth,0\fheight)},
legend cell align={left},
legend style={font=\scriptsize,at={(0.01,0.85)}, anchor=south west, draw=white!80.0!black},
tick align=outside,
tick pos=left,
x grid style={white!69.01960784313725!black},
xmajorgrids,
xmin=0, xmax=840,
xtick style={color=black},
y grid style={white!69.01960784313725!black},
ymajorgrids,
ymin=0, ymax=810,
ytick style={color=black},
hide y axis,
hide x axis,
legend columns=2,
]

\addplot [solid, semithick, color0]
table [row sep=crcr] {%
-1 -1\\
-2 -2\\
};
\addlegendentry{BW = 100.0 MHz}

\addplot [dashed, semithick, color0]
table [row sep=crcr] {%
-1 -1\\
-2 -2\\
};
\addlegendentry{BW = 400.0 MHz}

\end{axis}

\begin{axis}[
width=0.951\fwidth,
height=\fheight,
at={(0\fwidth,0\fheight)},
legend cell align={left},
legend style={font=\scriptsize,at={(0.01,1.02)}, anchor=south west, draw=white!80.0!black},
tick align=outside,
tick pos=left,
x grid style={white!69.01960784313725!black},
xmajorgrids,
xmin=0, xmax=840,
xtick style={color=black},
y grid style={white!69.01960784313725!black},
ymajorgrids,
ymin=0, ymax=810,
ytick style={color=black},
hide y axis,
hide x axis,
legend columns=3,
]

\addplot [solid, semithick, mark=*, color0]
table [row sep=crcr] {%
-1 -1\\
-2 -2\\
};
\addlegendentry{$d=50$ m}

\addplot [solid, semithick, mark=o, color1]
table [row sep=crcr] {%
-1 -1\\
-2 -2\\
};
\addlegendentry{$d=150$ m}

\addplot [solid, semithick, mark=diamond, color4]
table [row sep=crcr] {%
-1 -1\\
-2 -2\\
};
\addlegendentry{$d=250$ m}

\end{axis}

\end{tikzpicture}
	\caption{Throughput}
	\label{fig:thr}
\end{subfigure}%
\hfill%
\begin{subfigure}{0.48\textwidth}
	\centering
	\setlength\fwidth{\columnwidth}
	\setlength\fheight{.5\columnwidth}
	% This file was created by matplotlib2tikz v0.7.4.
\begin{tikzpicture}

\definecolor{color0}{rgb}{0.12156862745098,0.466666666666667,0.705882352941177}
\definecolor{color1}{rgb}{1,0.498039215686275,0.0549019607843137}
\definecolor{color2}{rgb}{0.172549019607843,0.627450980392157,0.172549019607843}
\definecolor{color3}{rgb}{0.83921568627451,0.152941176470588,0.156862745098039}
\definecolor{color4}{rgb}{0.580392156862745,0.403921568627451,0.741176470588235}
\definecolor{color5}{rgb}{0.549019607843137,0.337254901960784,0.294117647058824}
\pgfplotsset{every tick label/.append style={font=\scriptsize}}

\begin{axis}[
width=0.951\fwidth,
height=\fheight,
at={(0\fwidth,0\fheight)},
legend cell align={left},
legend style={at={(0.03,0.97)}, anchor=north west, draw=white!80.0!black},
tick align=outside,
tick pos=left,
x grid style={white!69.01960784313725!black},
xlabel={Source Rate [Mbps]},
xmajorgrids,
xmin=0, xmax=840,
xtick style={color=black},
y grid style={white!69.01960784313725!black},
ylabel={Average Latency [ms]},
ymajorgrids,
ymin=0, ymax=40,
ytick style={color=black},
ylabel style={font=\footnotesize\color{white!15!black}},
xlabel style={font=\footnotesize\color{white!15!black}}
]
\addplot [semithick, mark=*, color0]
table [row sep=crcr] {%
819.2 22.9135189956797\\
409.6 23.3189555977889\\
182.044444444444 2.34604603740566\\
81.92 1.20522329151833\\
8.192 0.28232517290461\\
};
% \addlegendentry{BW=100.0MHz - d=50.0 m}

\addplot [semithick, mark=*, mark options={solid,}, color0, dashed]
table [row sep=crcr] {%
819.2 1.49756589748221\\
409.6 1.22180299152167\\
182.044444444444 1.50657742269421\\
81.92 1.6251821950362\\
8.192 1.01113999477658\\
};
% \addlegendentry{BW=400.0MHz - d=50.0 m}

\addplot [semithick, mark=o, color1]
table [row sep=crcr] {%
819.2 25.9004746347383\\
409.6 27.6565524511706\\
182.044444444444 7.43689703997293\\
81.92 4.36565945980567\\
8.192 2.57000121428546\\
};
% \addlegendentry{BW=100.0MHz - d=150.0 m}

\addplot [semithick, mark=o, mark options={solid,}, color1, dashed]
table [row sep=crcr] {%
819.2 5.27905112757609\\
409.6 4.21124148130884\\
182.044444444444 3.96869708501577\\
81.92 2.90191251006789\\
8.192 2.76663856120835\\
};
% \addlegendentry{BW=400.0MHz - d=150.0 m}

\addplot [semithick, mark=diamond, color4]
table [row sep=crcr] {%
819.2 34.8118691718034\\
409.6 32.3625334995671\\
182.044444444444 22.9389580139467\\
81.92 7.38078353732669\\
8.192 4.19684751960863\\
};
% \addlegendentry{BW=100.0MHz - d=250.0 m}

\addplot [semithick, mark=diamond, mark options={solid,}, color4, dashed]
table [row sep=crcr] {%
819.2 11.6895467056911\\
409.6 9.5103028900958\\
182.044444444444 5.78748168266766\\
81.92 6.17122302479936\\
8.192 4.42041017152716\\
};
% \addlegendentry{BW=400.0MHz - d=250.0 m}

\end{axis}

\begin{axis}[
width=0.951\fwidth,
height=\fheight,
at={(0\fwidth,0\fheight)},
legend cell align={left},
legend style={font=\scriptsize,at={(0.01,0.85)}, anchor=south west, draw=white!80.0!black},
tick align=outside,
tick pos=left,
x grid style={white!69.01960784313725!black},
xmajorgrids,
xmin=0, xmax=840,
xtick style={color=black},
y grid style={white!69.01960784313725!black},
ymajorgrids,
ymin=0, ymax=765.634092886124,
ytick style={color=black},
hide y axis,
hide x axis,
legend columns=2,
]

\addplot [solid, semithick, color0]
table [row sep=crcr] {%
-1 -1\\
-2 -2\\
};
\addlegendentry{BW = 100.0 MHz}

\addplot [dashed, semithick, color0]
table [row sep=crcr] {%
-1 -1\\
-2 -2\\
};
\addlegendentry{BW = 400.0 MHz}

\end{axis}

\begin{axis}[
width=0.951\fwidth,
height=\fheight,
at={(0\fwidth,0\fheight)},
legend cell align={left},
legend style={font=\scriptsize,at={(0.01,1.02)}, anchor=south west, draw=white!80.0!black},
tick align=outside,
tick pos=left,
x grid style={white!69.01960784313725!black},
xmajorgrids,
xmin=0, xmax=840,
xtick style={color=black},
y grid style={white!69.01960784313725!black},
ymajorgrids,
ymin=0, ymax=765.634092886124,
ytick style={color=black},
hide y axis,
hide x axis,
legend columns=3,
]

\addplot [solid, semithick, mark=*, color0]
table [row sep=crcr] {%
-1 -1\\
-2 -2\\
};
\addlegendentry{$d=50$ m}

\addplot [solid, semithick, mark=o, color1]
table [row sep=crcr] {%
-1 -1\\
-2 -2\\
};
\addlegendentry{$d=150$ m}

\addplot [solid, semithick, mark=diamond, color4]
table [row sep=crcr] {%
-1 -1\\
-2 -2\\
};
\addlegendentry{$d=250$ m}

\end{axis}

\end{tikzpicture}
	\caption{Latency}
	\label{fig:delay}
\end{subfigure}
\caption{Metrics for \texttt{vehicular-simple-one.cc}, \texttt{V2V-Highway} scenario, for different source rates, bandwidth BW and distance between vehicles $d$.}
\label{fig:sinr}
\end{figure*}
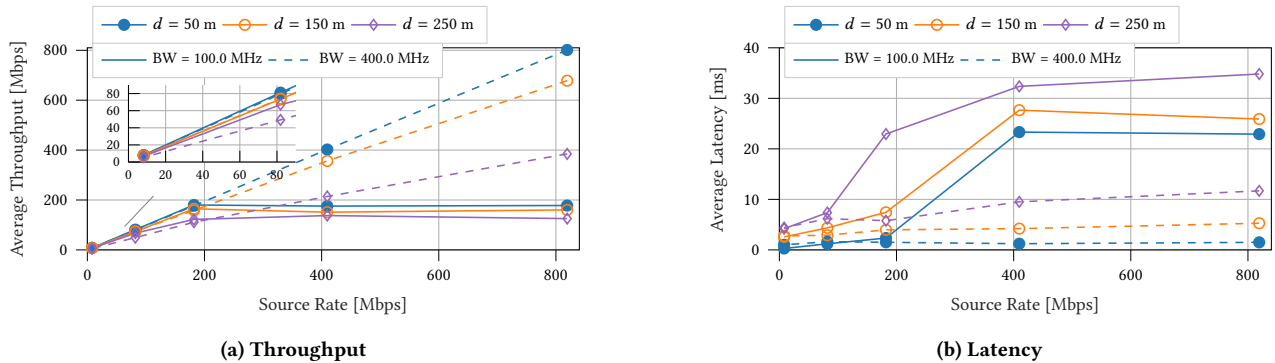

The \texttt{MmWaveVehicularHelper} also configures and connects to  each device another helper, called \texttt{MmWaveVehicularTracesHelper}, which, at runtime, generates a trace of the \gls{sinr} and \gls{mcs} for each transmitted packet.

Finally, \texttt{PairDevices} configures a bearer and connects, pair by pair, all the devices in a container of \texttt{NetDevices} passed as input argument. This makes it possible to create multiple groups of vehicles, with independent scheduling patterns, that generate interference with concurrent transmissions. The vehicles in the same group are all logically connected, thus packets could (in principle) be exchanged among any pair of nodes.

\subsection{Unit Test}
\label{sub:test}
The unit test suite contains four tests. \texttt{MmWaveVehicularSidelink\-SpectrumPhyTestSuite} has a test case that checks if the \gls{snr} computed by \texttt{MmWaveSidelinkSpectrumPhy} for a single transmission is in line with the expected \gls{snr} (considering isotropic antennas and an ideal channel). Similarly, \texttt{MmWaveVehicularInterferenceTestSuite} tests if the interference among two groups of vehicles is correctly computed.
\texttt{MmWaveVehicularRateTestCase}, instead, features vehicles equipped with a full protocol stack (with \gls{udp} at the transport layer), and tests full buffer transmissions for different values of the \gls{mcs} (i.e., from 0 to 28), checking if there is a one-to-one correspondence between transmitted and received packets.

\section{Example Scenarios}
\label{sec:scenarios}

We provide four examples of scenarios in the \texttt{examples} folder. They represent different simple use cases for the module, and can be used as a reference by a user of the module to understand which are the necessary steps to generate a vehicular scenario at mmWave frequencies. In particular, these are the common operations to be implemented:
\begin{compactenum}
	\item Configure relevant parameters for the simulation scenario, using the ns-3 \texttt{Config} attribute system. For example, relevant parameters are related to the carrier frequency, the system bandwidth, whether to use \gls{amc} or not, the type of \gls{rlc} and the size of its buffer;
	\item Create nodes of the scenario, setting their positions and mobility models;
	\item Initialize the \texttt{MmWaveHelper} object, for which it is possible to set the numerology and the channel models, as discussed in Sec.~\ref{sec:helpers};
	\item Install the \texttt{MmWaveVehicularNetDevice} objects in the nodes
	\item Configure the TCP/IP stack and assign IP addresses to the vehicles
	\item Configure and install the applications.
\end{compactenum}

The simplest scenario is \texttt{vehicular-simple-one.cc}. In this, there are two devices in the same lane, one behind the other at fixed distance and speed. It is possible to select the speed and the simulation duration from the command line, as well as the \gls{mcs} value or if \gls{amc} should be used. The applications are an echo client and server using \gls{udp} as transport. The \texttt{vehicular-simple-two.cc} example, instead, features two groups with two vehicles each, moving in two different lanes at opposite speeds. Each vehicle exchanges \gls{udp} packets with that in the same lane.
\texttt{vehicular-simple-three.cc}, instead, also has four vehicles, grouped in pairs of two, but moving as a platoon, on the same lane, at a constant speed and at a safety distance of 20 m. Finally, \texttt{vehicular-simple-four.cc} is a scenario with three vehicles, with two of them moving at a constant distance and speed, and the third moving away from the first two. One of the vehicles acts as a server, which echoes the \gls{udp} packets sent by the others.

As an example, we report in this Section some metrics that it is possible to obtain by running the available examples. Figs~\ref{fig:thr} and~\ref{fig:delay} have been obtained choosing different configurations of parameters in \texttt{vehicular-simple-one.cc}, such as the initial distance $d$ between the vehicles, the bandwidth (100 MHz or 400 MHz, which is the maximum for an NR carrier), and the source rate at the application layer. In this simulation campaign the channel was configured to transmit using a carrier frequency $f_c = 28$~GHz, with the \gls{los} condition computed using the equations outlined in Sec.~\ref{sec:channel} for the reference scenario \texttt{V2V-Highway}. Moreover, the \texttt{LteRlcUm} buffer size was set to $500$ packets.
Buffering is necessary to guarantee high throughput, as (i) the scheduling at the \gls{mac} layer happens on a slot basis (i.e., 250 $\mu$s with numerology 2), but the application may generate packets more regularly, with interpacket intervals of up to 10 $\mu$s; (ii) the default allocation pattern is used, i.e., each vehicle is assigned a single slot per subframe; and (iii) the fluctuations of the channel may temporarily degrade the capacity available at the physical layer. On the other hand, an excessively large buffer could lead to an increase in the experienced delay of the communication, due to packets buffering in the queue.

\begin{figure*}
  \setlength{\belowcaptionskip}{-0.33cm}
	\begin{subfigure}{0.48\textwidth}
	\centering
	\setlength\fwidth{\columnwidth}
	\setlength\fheight{.5\columnwidth}
	\input{figures/sinr-highway.tex}
	\caption{\texttt{V2V-Highway} scenario}
	\label{fig:sinr-hw}
\end{subfigure}%
\hfill%
\begin{subfigure}{0.48\textwidth}
	\centering
	\setlength\fwidth{\columnwidth}
	\setlength\fheight{.5\columnwidth}
	\input{figures/sinr-urban.tex}
	\caption{Urban scenario}
	\label{fig:sinr-un}
\end{subfigure}
\caption{SINR for a pair of vehicles in \texttt{vehicular-simple-two.cc}. $4\times4$ ($8\times8$) indicates that both endpoints have antenna arrays with $4$ ($8$) elements.}
\label{fig:sinr}
\end{figure*}

Fig.~\ref{fig:thr} compares the throughput (averaged over the simulation time of 2 s, and over 100 independent simulation runs) for different values of the source rate. It can be seen that both configurations with 100 MHz and 400 MHz sustain a source rate of up to 80 Mbps, with small differences when varying $d$, but the performance gap becomes more marked for higher source rates. Fig.~\ref{fig:delay}, instead, reports the average delay in the same conditions. It has to be highlighted that the end-to-end delay obtained when the channel bandwidth is set to $400$~MHz, the initial distance is $50$~m and the source rate is $\sim8$ Mbps, is in the order of microseconds ($\sim1$ \gls{ofdm} symbol). This is motivated by the fact that the traffic injected in the link is negligible with respect to the available resources, and the packet only experiences transmission delay. Future extension of this work will study how to model the amount of delay introduced by the processing of packets at the different layers of the protocol stack.
%In both figures, each point represents the mean value obtained from 50 different runs of the same set-up.

Fig.~\ref{fig:sinr} shows the trend of the \gls{sinr} over time, for a single run of \texttt{vehicle-simple-two.cc}, to validate how the interference is modeled by \millicar{}. The plots show a decrease in the received \gls{sinr}, with a minimum at $\sim1$ s, and a consequent increase. This reduction is caused by a separate pair of vehicles interfering with the transmission that the nodes considered for the \gls{sinr} metric are carrying out. This setup has been evaluated in both scenarios defined by 3GPP, i.e. \texttt{V2V-Highway} (Fig.~\ref{fig:sinr-hw}) and \texttt{V2V-Urban} (Fig.~\ref{fig:sinr-un}), and with different dimensions of the antenna array. Notice that, while different architectures have a significant impact on the received power and, thus, on the \gls{sinr}, the two scenarios differ only by 1.2 dB (for the configuration with 4 antennas at the transmitter and 4 at the receiver).

\section{Conclusions}
\label{sec:conclusions}

The wireless networking standardization bodies have started focusing on new market verticals to find new use cases and applications for 5G and beyond. For example, the 3GPP is considering vehicle-to-vehicle communications in NR V2X, which will support vehicular networking in the mmWave frequency bands. In this paper, we introduced \millicar{}, the first implementation of an open-source ns-3 module for the simulation of NR-V2X networks at \glspl{mmwave}. The module enables end-to-end, full-stack simulations of vehicular networks with a 3GPP channel model for \gls{v2v} propagation and fading at mmWaves, physical and \gls{mac} layers redesigned for NR V2X, and integration with the higher layers of the protocol stack (e.g., \gls{rlc}) from ns-3.
We believe that this contribution, being an open-source tool, easily extensible and available to the overall wireless community, will help foster more research on the design and performance evaluation of \gls{mmwave} vehicular networks.

As future works, we plan to improve the functionalities of \millicar{} by adding a number of new features, including a more robust and realistic beam management framework and a multiple medium access scheme, and by keeping the module updated to the latest proposals in the 3GPP standardization process.

% \vspace{-.1cm}
% \begin{acks}
% 	This work was partially supported  by the U.S. Commerce Department/NIST through the project ``An End-to-End Research Platform for Public Safety Communications above 6 GHz''.
% \end{acks}

\bibliographystyle{ACM-Reference-Format.bst}
\bibliography{../../bibl.bib}

% \balancecolu
\end{document}